\documentclass[aps,twocolumn]{revtex4}
\usepackage{bm}
\usepackage{amsmath}
\usepackage{graphicx}
\usepackage{subfigure}
\usepackage[usenames,dvipsnames]{color}
\definecolor{darkblue}{RGB}{0,0,196}
\usepackage[colorlinks=true,linkcolor=darkblue,citecolor=darkblue,urlcolor=darkblue]{hyperref}
\usepackage{setspace}
\usepackage{hyperref}
\usepackage{xcolor}
\hypersetup{
  colorlinks   = true, 
  urlcolor     = red, 
  linkcolor    = blue, 
  citecolor   = blue 
}
\usepackage{footmisc}
\usepackage[makeroom]{cancel}
\usepackage{comment}
\usepackage{lineno}
\def\be{\begin{equation}}
\def\ee{\end{equation}}
\def\ba{\begin{eqnarray}}
\def\ea{\end{eqnarray}}
\usepackage{graphicx}
\usepackage{amsmath,bbm}
\usepackage{amssymb,bm}
\usepackage{yfonts}

\begin{document}
\title{Study of Transverse Spherocity and Azimuthal Anisotropy in Pb-Pb collisions at $\sqrt{s_{\rm{NN}}} = 5.02$~TeV using A Multi-Phase Transport Model}
\author{Neelkamal Mallick}
\author{Raghunath Sahoo\footnote{Corresponding author: $Raghunath.Sahoo@cern.ch$}}
\affiliation{Department of Physics, Indian Institute of Technology Indore, Simrol, Indore 453552, India}
\author{Sushanta Tripathy}
\author{Antonio Ortiz}
\affiliation{Instituto de Ciencias Nucleares, Universidad Nacional Aut\'onoma de M\'exico, M\'exico Distrito Federal 04510, M\'exico }

\begin{abstract}
\noindent
Transverse spherocity is an event shape observable having a unique capability to separate the events based on their geometrical shapes. Recent results from experiments at the LHC suggest that transverse spherocity is an important event classifier in small collision systems. In this work, we use transverse spherocity for the first time in heavy-ion collisions and perform an extensive study on azimuthal anisotropy of charged particles produced in Pb-Pb collisions at $\sqrt{s_{\rm{NN}}} = 5.02$~TeV using A Multi-Phase Transport Model (AMPT). The azimuthal anisotropy is estimated using the 2-particle correlation method, which suppresses the non-flow effects significantly with an appropriate pseudorapidity gap of particle pairs. The results from AMPT are compared with estimations from PYTHIA8 (Angantyr) model and it is found that with the chosen pseudorapidity gap, the residual non-flow effects become negligible. We found that the high spherocity events have nearly zero elliptic flow while low spherocity events contribute significantly to elliptic flow of spherocity-integrated events. 

\end{abstract}
\date{\today}
\maketitle 

\section{Introduction}
\label{intro}
Quark Gluon Plasma (QGP), a deconfined state of quarks and gluons, is believed to be produced in ultra-relativistic heavy-ion collisions at the Large Hadron Collider (LHC) at CERN, Switzerland and Relativistic heavy-ion collider (RHIC) at BNL, USA. However, due to its very short lifetime we do not have any direct evidence of possible QGP formation, instead several indirect signatures such as strangeness enhancement, direct photon measurements, elliptic flow, suppression of charmonia etc. suggest that the formation of QGP is highly probable in such collisions. Traditionally, the results from collisions of protons are considered as a baseline to understand
the nuclear medium formation and its characterization in heavy-ion collisions~\cite{Busza:2018rrf}. Recent measurements of heavy-ion-like behaviours such as ridge-like structures~\cite{Khachatryan:2016txc} and strangeness enhancement~\cite{ALICE:2017jyt} in $pp$ collisions at the LHC, have created concerns in the heavy-ion physics community. To understand the dynamics of small collision systems, an event shape observable, transverse spherocity, has been introduced recently~\cite{Cuautle:2014yda,Cuautle:2015kra,Ortiz:2017jho,Salam:2009jx,Bencedi:2018ctm,Banfi:2010xy,Khuntia:2018qox,Tripathy:2019blo}. From these studies, it was observed that transverse spherocity has unique capability to separate the events based on their geometrical shapes, i.e. jetty and isotropic in small collision systems. Also, the preliminary measurement by ALICE collaboration at the LHC suggests that using transverse spherocity one can increase/decrease the enhanced production of strangeness in high-multiplicity $pp$ collisions by separating the event types~\cite{Tripathy:2020jue,Tripathy:2020lla,adrianQM}. After its successful implementation in small collision systems~\cite{Acharya:2019mzb}, the use of transverse spherocity in heavy-ion collisions may reveal new and unique results from heavy-ion collisions where the production of a QGP medium is already established. Event shape study based on transverse spherocity in heavy-ion collisions will also complement the current event shape approach based on flow vector in LHC experiments~\cite{Aad:2015lwa,Poskanzer:1998yz}.

Studies of the azimuthal anisotropy of particle production via anisotropic flow contribute significantly to the characterization of the system formed in heavy-ion collisions~\cite{Voloshin:2008dg,Ollitrault:1992bk}. Presence of finite anisotropic flow at the LHC is ascribed primarily to the response of produced QGP to fluctuations of the initial energy density profile of the colliding nucleons. Elliptic flow measures the momentum anisotropy of the final state particles, which is sensitive to the initial geometry of the overlap region and to the transport properties and equation of state of the system. It is defined as the second-order Fourier coefficient of the particle azimuthal distribution~\cite{v2}. Based on the anisotropic flow studies, it has been shown that the shear viscosity to entropy density ratio of the QGP produced in ultra-relativistic heavy-ion collisions at RHIC and LHC has a value close to 1/4$\pi$, which is the lower bound obtained in strong-coupling calculations based on the AdS/CFT conjecture~\cite{Kovtun:2004de}. This study has revealed QGP
as a {\it ``perfect fluid"} found in nature \cite{Adams:2005dq}.

In this work, we use transverse spherocity for the first time in heavy-ion collisions and perform an extensive study on azimuthal anisotropy of charged particles produced in Pb-Pb collisions at $\sqrt{s_{\rm{NN}}} = 5.02$~TeV using A Multi-Phase Transport Model (AMPT)~\cite{AMPT2}. The purpose of the present analysis is to introduce a multi-differential method to study the dynamics of heavy-ion collisions using transverse spherocity and collision centrality for the first time. Thus, we have used the default parameterization available in AMPT model. Also, for a cross-check, the results of elliptic flow are compared with PYTHIA8 (Angantyr) model~\cite{Bierlich:2018xfw}, where one does not expect finite elliptic flow after removal of non-flow effects. We believe that the present analysis would motivate experimentalists to pursue such a novel method in experiments at RHIC and the LHC. To match with the experimental data, one can vary the tunes of the AMPT model, which is out of the scope of this manuscript. However, we also show a comparison of the results of default settings from string melting mode of AMPT with the experimental data, wherever possible.

The paper is organised as follows. We begin with a brief introduction and motivation for the study in Section~\ref{intro}. In Section~\ref{section2}, the detailed analysis methodology along with brief description about AMPT and PYTHIA (Angantyr) are given. Section~\ref{section3} discusses about the method of calculation of elliptic flow and  discussion on results. Finally the results are summarized in Section~\ref{section4}.

\section{Event Generation and Analysis Methodology}
\label{section2}

In this section, we begin with a brief introduction on the event generators used in this analysis. Then, we proceed to define the transverse spherocity before performing a detailed analysis.

\subsection{A Multi-Phase Transport (AMPT) model}
\label{formalism}
A Multi-Phase Transport Model contains four components namely, initialization of collisions, parton transport after initialization, hadronization mechanism and hadron transport~\cite{AMPT2}.  The initialization of the model is done using HIJING~\cite{ampthijing}, which calculates the differential  cross-section of the produced mini-jets in $pp$ collisions. Then, the produced partons calculated in $pp$ collisions is converted into A-A and $p$-A collisions by incorporating parametrized shadowing function and nuclear overlap function using inbuilt Glauber model within HIJING. Similarly, initial low-momentum partons are produced from parametrized coloured string fragmentation mechanisms. Initial low-momentum partons are separated from high momentum partons by a momentum cut-off. The produced particles are initiated into parton transport part, Zhang’s Parton Cascade (ZPC) model~\cite{amptzpc}. In the String Melting version of AMPT (AMPT-SM), melting of colored strings into low momentum partons take place at the start of the ZPC. It is calculated using Lund FRITIOF model of HIJING. The resulting partons undergo multiple scatterings which take place when any two partons are within distance of minimum approach. In AMPT-SM, the transported partons are finally hadronized using spatial coalescence mechanism~\cite{Lin:2001zk,He:2017tla}. The produced hadrons further undergo final evolution in a relativistic transport mechanism~\cite{amptart1, amptart2} via meson-meson, meson-baryon and baryon-baryon interactions. There is also a default version of AMPT, where instead of coalescing the partons, fragmentation mechanism using Lund fragmentation parameters $a$ and $b$ are used for hadronizing the transported partons. However, the particle flow and spectra at the mid-$p_{\rm T}$ regions are well explained by quark coalescence mechanism for hadronization~\cite{ampthadron1,ampthadron2,ampthadron3}. Thus, we have used AMPT-SM mode (AMPT version 2.26t7) for all of our calculations in our current work. The AMPT settings in the current work, are same as reported in Ref.~\cite{Tripathy:2018bib} and they are set to default settings of AMPT-SM mode. For the input of impact parameter values for different centralities in Pb-Pb collisions, we have used Ref.~\cite{Loizides:2017ack}. One should note here that, high central collisions correspond to low impact parameter values and higher final state charged-particle multiplicity ($\langle dN_{\rm ch}/d\eta \rangle$). 

\subsection{PYTHIA8 (Angantyr)}
PYTHIA8~\cite{Sjostrand:2014zea} is a parton-based event generator, which is a widely known Monte Carlo generator for high-energy collider physics with an emphasis on $pp$ collisions. It is incorporated with  several known  physics  mechanisms for $e^{+}e^{-}$, $pp$ and $p\bar{p}$ collisions like hard and soft interactions, parton distributions, initial- and final-state parton showers, multipartonic interactions (MPI), string fragmentation, color reconnection and resonance decays. The main event of a $pp$ collision in PYTHIA is represented by 2-to-2 matrix elements with hard parton scatterings (defined at the leading order) and it is complemented by the leading-logarithmic approximation of parton showers. It also contains the underlying event, which includes beam-remnants, MPI, initial state and final state radiation. The  hadronization is done using the Lund string fragmentation model and in the Color Reconnection (CR) picture of the model, the strings between partons can be rearranged.

In the latest versions, PYTHIA8 includes Angantyr model for the prediction of results in heavy-ion collisions. The main idea of Angantyr in PYTHIA is to extrapolate the $pp$ dynamics, as described by the model for MPIs and underlying events in the PYTHIA8, to heavy-ion collisions, retaining as much information as possible from $pp$ collisions~\cite{Bierlich:2018xfw}. In order to make predictions for heavy-ion collisions, different components of a standard PYTHIA8 simulation was modified and it was tuned with the results from $e^{+}e^{-}$, $pp$ and $p\bar{p}$ collisions, which is discussed in Ref.~\cite{Bierlich:2018xfw}. Thus, the current model retains the production mechanisms from small collision systems. So, it is evident that Angantyr does not include an assumption of a hot thermalised medium. Therefore, it can serve as a baseline for understanding the non-collective background to observables sensitive to collective behavior. We shall discuss about the comparison of the two particle elliptic flow coefficient in minimum bias Pb-Pb collisions at $\sqrt{s_{\rm{NN}}} = 5.02$~TeV from AMPT and PYTHIA8 (Angantyr) in Section~\ref{section3}. We have used the default settings of PYTHIA version 8.235~\cite{Bierlich:2018xfw} in our current work.

\begin{figure}[ht]
\includegraphics[scale=0.42]{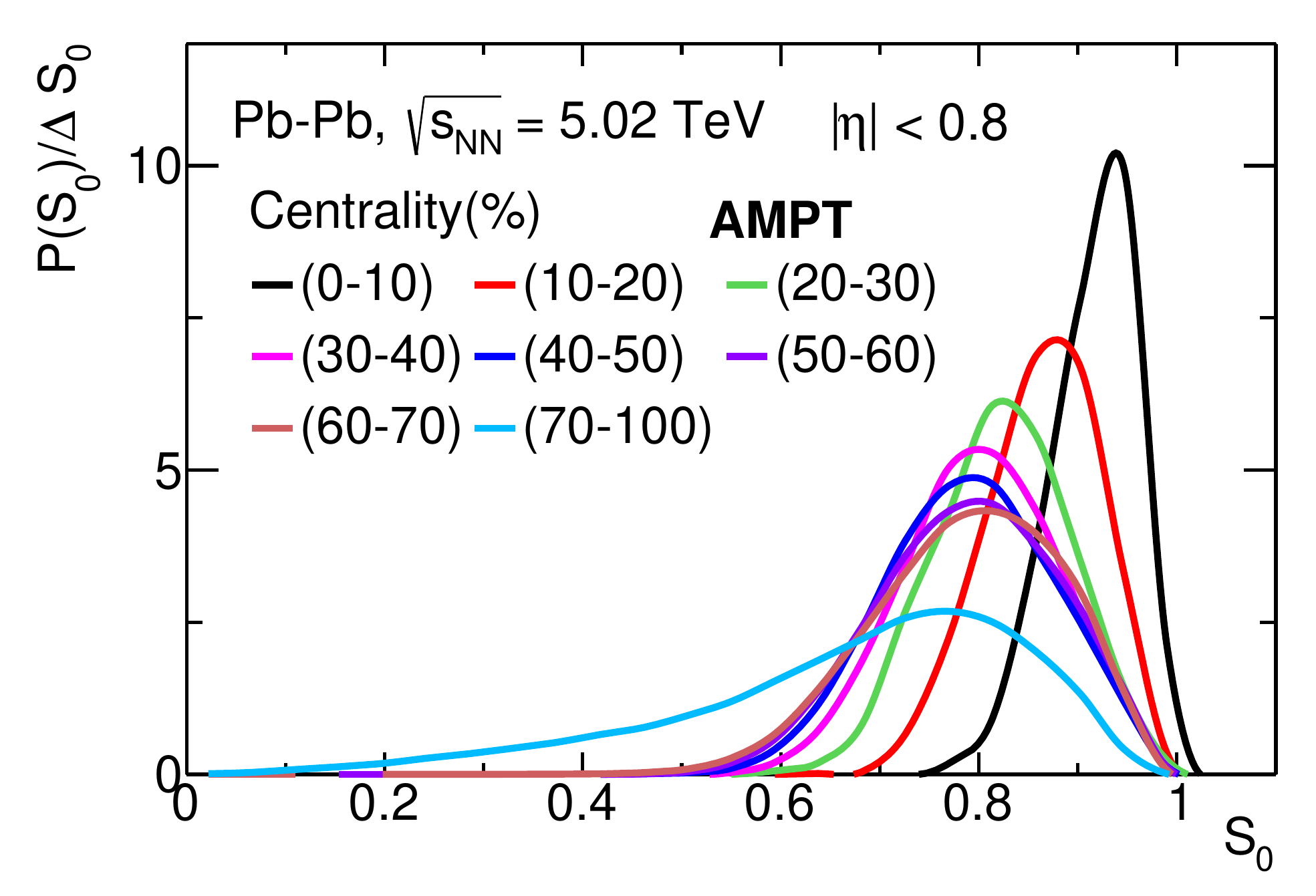}
\caption[]{(Color Online) Spherocity distributions for different centrality classes in Pb-Pb collisions at $\sqrt{s_{\rm{NN}}} = 5.02$~TeV.}
\label{sp_common}
\end{figure}

\subsection{Transverse Spherocity}

Transverse spherocity is an event property which is defined for a unit vector $\hat{n} (n_{T},0)$ that minimizes the ratio~\cite{Cuautle:2014yda, Cuautle:2015kra}:
\begin{eqnarray}
S_{0} = \frac{\pi^{2}}{4} \bigg(\frac{\Sigma_{i}~|\vec p_{T_{i}}\times\hat{n}|}{\Sigma_{i}~p_{T_{i}}}\bigg)^{2}.
\label{eq1}
\end{eqnarray}

By restricting it to the transverse plane, transverse spherocity becomes infrared and collinear safe~\cite{Salam:2009jx}. By construction, the extreme limits of transverse spherocity are related to specific configurations of events in transverse plane. The value of transverse spherocity ranges from 0 to 1, which is ensured by multiplying the normalization constant $\pi{^2}/4$ in Eq.~\ref{eq1}. In $pp$ collisions~\cite{Acharya:2019mzb}, transverse spherocity becoming 0 means, the events are pencil-like (back-to-back structure) and called as jetty events, while 1 would mean the events are isotropic. The jetty events are usually the hard events while the isotropic ones are the result of soft processes. 

Here onwards, for the sake of simplicity the transverse spherocity is referred as spherocity. To disentangle the low and high spherocity events from the average-shaped events, we have applied spherocity cuts on our generated events. The spherocity distributions are selected in the pseudo-rapidity range of $|\eta|<0.8$ with a minimum constraint of 5 charged particles with $p_{\rm{T}}>$~0.15~GeV/$c$ to recreate the similar conditions as in ALICE experiment at the LHC. The spherocity distributions for different centrality classes for Pb-Pb collisions at $\sqrt{s_{\rm{NN}}} = 5.02$~TeV are shown in Fig.~\ref{sp_common}. We observe that the spherocity distributions are shifted towards one from peripheral to central collisions. This effect is an artefact of increasing final state charged-particle multiplicity from peripheral to central collisions. Low-$S_{0}$ and high-$S_{0}$ are chosen with 20\% cuts on spherocity distributions in left and right side, respectively in Pb-Pb collisions at $\sqrt{s_{\rm{NN}}} = 5.02$~TeV for different centrality classes. The limits for low-$S_{0}$ and high-$S_{0}$ can be seen in Table~\ref{tab:1}. 

\begin{table}[ht!]
\begin{center}
\caption{Low 20 \% and high 20\% cuts on spherocity distribution in Pb-Pb collisions at $\sqrt{s_{\rm{NN}}} = 5.02$~TeV for different centrality classes. }

\label{tab:1}
\begin{tabular}{ |p{2cm}|p{2cm}|p{2cm}|}
 \hline
 Centrality (\%) & Low-$S_{0}$ & High-$S_{0}$\\
\hline
0-10        		& 0 -- 0.880         &	0.953 -- 1 \\
10-20	         & 0 -- 0.813        &    0.914 -- 1 \\
20-30		& 0 -- 0.760	  &	0.882 -- 1 \\
30-40		& 0 -- 0.735	  &	0.869 -- 1 \\
40-50		& 0 -- 0.716	  &	0.865 -- 1 \\
50-60		& 0 -- 0.710	  &	0.870 -- 1 \\
60-70		& 0 -- 0.707	  &	0.873 -- 1 \\	
70-100		& 0 -- 0.535	  &	0.822 -- 1 \\
 \hline
 \end{tabular}
 \end{center}
\end{table}

\begin{figure*}[ht]
\includegraphics[scale=0.85]{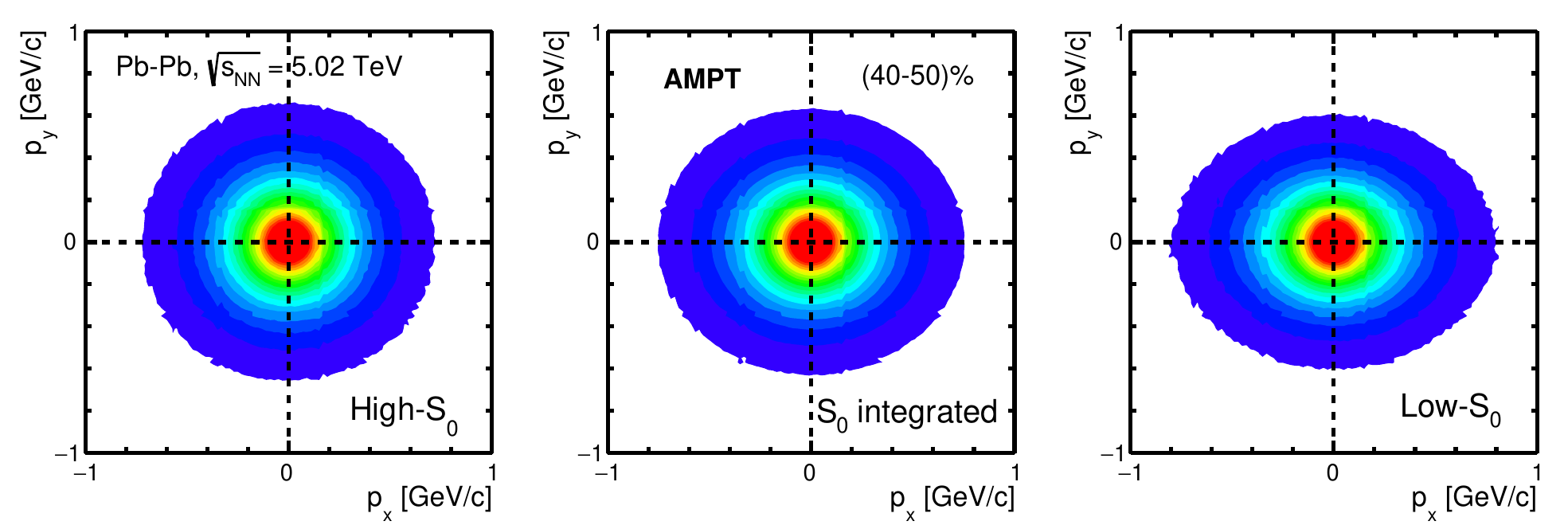}
\caption[]{(Color Online) Transverse momentum space correlation ($p_{\rm y}$ vs. $p_{\rm x}$) in different spherocity classes for (40-50)\% Pb-Pb collisions at $\sqrt{s_{\rm{NN}}} = 5.02$~TeV in AMPT model. Here, z-axis is the beam direction.}
\label{sp_mom}
\end{figure*}

Figure~\ref{sp_mom} shows the transverse momentum space correlation ($p_{\rm y}$ vs. $p_{\rm x}$) in different spherocity classes for (40-50)\% Pb-Pb collisions at $\sqrt{s_{\rm{NN}}} = 5.02$~TeV in AMPT model. The spherocity-integrated events clearly show the presence of spatial anisotropy in the initial stage due to the initial almond shape of nuclear overlap region in semi-central collisions. Due to high pressure gradient along x-axis, particles emerging along x-axis carry more momentum than y-axis ($p_{\rm x}$ $>$ $p_{\rm y}$). This indicates the presence of elliptic flow. Now, with different spherocity classes the transverse momentum space correlation can be varied as seen in Figure~\ref{sp_mom}. The events with high spherocity values show a circular transverse momentum space correlation implying almost zero azimuthal anisotropy. However, with varying the spherocity cuts from high to low, the transverse momentum space correlation becomes more elliptical, which implies the presence of finite elliptic flow in low spherocity event classes. This is a testimony that spherocity is a unique tool, which can be used as an event classifier to study the collective effects in heavy-ion collisions.

\section{Results and Discussions}
\label{section3}

The anisotropic flow of different order can be characterized by the coefficients ($v_n$), which are obtained from a Fourier expansion of the momentum distribution of the charged particles. It is given by,
\begin{eqnarray}
E\frac{d^3N}{dp^3}=\frac{d^2N}{2\pi p_{\rm T}dp_{\rm T}dy}\bigg(1+2\sum_{n=1}^\infty v_n \cos[n(\phi -\psi_n)]\bigg)\,.\nonumber\\
\label{eq2}
\end{eqnarray}
Here, $\phi$ is the azimuthal angle in the transverse momentum plane and $\psi_n$ is the $n^{\text{th}}$ harmonic event plane angle~\cite{v2eventplane}. 

Taking $n$ = 2 in Eq.~\ref{eq2} gives the second order harmonics in the expansion and its coefficient, $v_2$ is calculated to provide the measure of the elliptic flow or azimuthal anisotropy. Thus, $v_{2}$ is defined as:
 \begin{eqnarray}
v_{2} = \langle \cos(2(\phi - \psi_2))\rangle
\label{eq3}
\end{eqnarray}

Obtaining the event plane angle is nearly impossible in real experiments. Thus, to compare with experimental data, we use two-particle correlation analysis to calculate the elliptic flow as obtained in experiments~\cite{Adam:2016izf,Voloshin:2008dg,Aad:2015lwa,Aaboud:2016yar}. The two-particle correlation method has an added advantage as by construction with a proper pseudo-rapidity cut, it would remove the residual non-flow effects in the elliptic flow. Non-flow effects are azimuthal correlations usually arise from jets and resonance decays, which are not associated with the symmetry planes.

The related observable for studying the properties of the medium is the correlation function between two particles in relative pseudorapidity ($\Delta\eta = \eta_a - \eta_b$) and azimuthal angle ($\Delta\phi = \phi_a - \phi_b$). The labels $a$ and $b$ denote the two particles in the pair, which are selected from different transverse momentum intervals. The two particle correlation function ($C(\Delta\eta ,\Delta\phi)$) can be constructed as the ratio of distributions for same-event pairs ($S(\Delta\eta ,\Delta\phi)$) and mixed-event pairs ($B(\Delta\eta ,\Delta\phi)$), given by:
\begin{eqnarray}
C(\Delta\eta ,\Delta\phi) = \frac{S(\Delta\eta ,\Delta\phi)}{B(\Delta\eta ,\Delta\phi)}
\label{eq4}
\end{eqnarray}

Here, mixed-event background ensures there is no non-uniformity and improves pair acceptance. For mixed-event background, five events are randomly chosen, hence it contains no physical correlations. As the primary aim is to obtain the anisotropy in azimuthal direction, this analysis focuses mainly on the shape of the correlation function in $\Delta\phi$. Thus the 1D correlation function for $\Delta\phi$ is given as;
\begin{eqnarray}
C(\Delta\phi) = \frac{dN_{\rm pairs}}{d\Delta\phi} = A \times \frac{\int S(\Delta\eta ,\Delta\phi) d\Delta\eta}{\int B(\Delta\eta ,\Delta\phi)d\Delta\eta}.
\label{eq5}
\end{eqnarray}
Here, the normalization constant ($A$) is given as $N_{\rm pairs}^{\rm mixed}/N_{\rm pairs}^{\rm same}$. $N_{\rm pairs}^{\rm mixed}$ and $N_{\rm pairs}^{\rm same}$ are the number of mixed-event pairs and same-event pairs, respectively in a chosen pseudorapidity gap of particle pairs used in the analysis. This scaling is applied to ensure that the number of pairs will be similar for same-events ($S$) and mixed-events background ($B$).

\begin{figure}[ht!]
\centering
\includegraphics[scale=0.42]{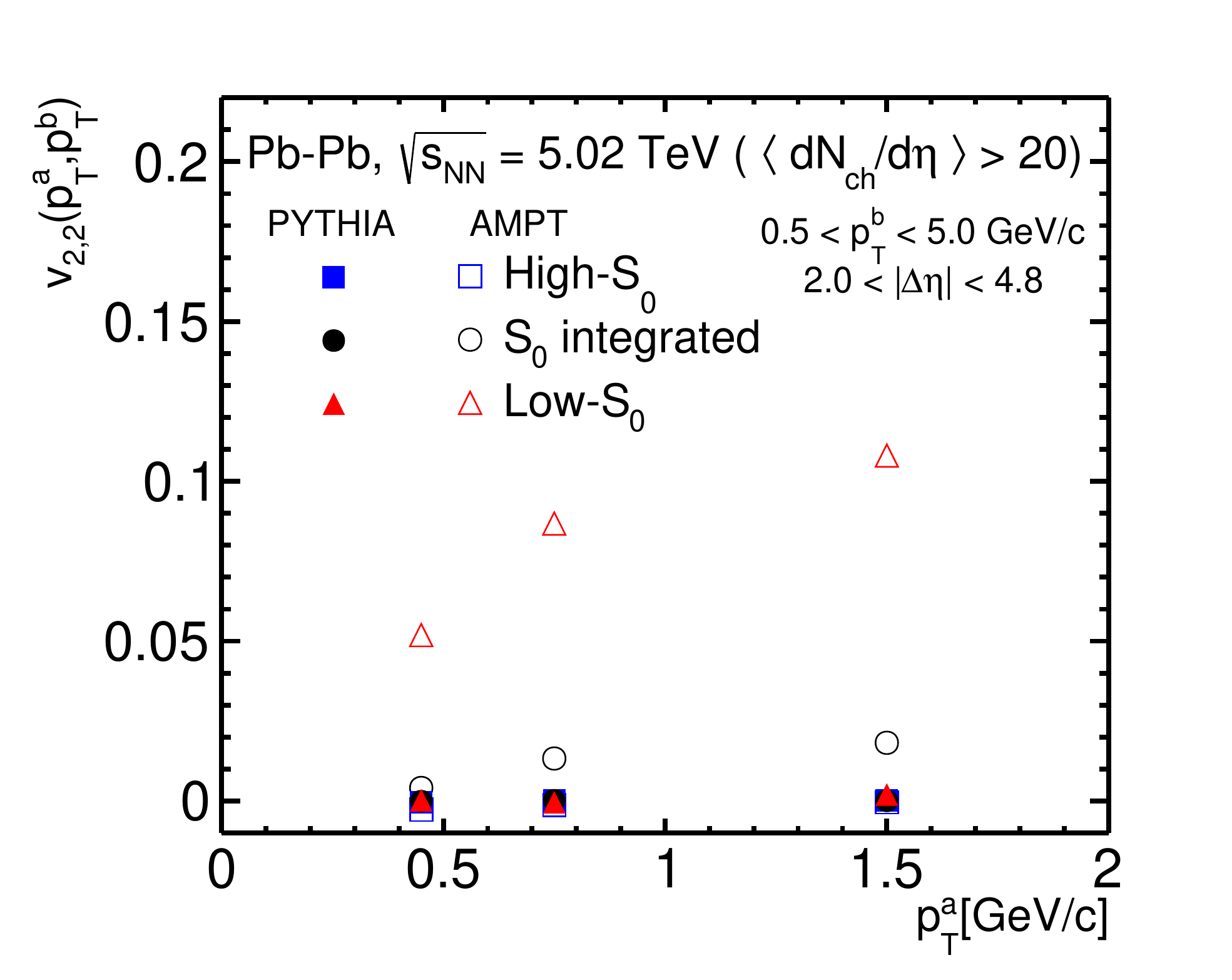}
\caption[width=18cm]{(Color Online) Two particle elliptic flow co-efficient ($v_{2,2}(p_T^a,p_T^b)$) as a function of $p_T^a$ for low-$S_{0}$ (red triangles), high-$S_{0}$ (blue squares) and spherocity-integrated (black circles) events in minimum bias Pb-Pb collisions at $\sqrt{s_{\rm NN}} = 5.02$ TeV in PYTHIA (full markers) and AMPT (empty markers). For the comparison and to have enough participants, the events are chosen with more than 20 charged-particle multiplicities in mid-rapidity ($|\eta| < 0.8$).}
\label{v22_modelcomp}
\end{figure}

Now, the pair distribution ($N_{\rm pairs}$) can be expanded in $\Delta\phi$ into a Fourier series:
\begin{eqnarray}
\frac{dN_{\rm pairs}}{d\Delta\phi} \propto 1+2\sum_{n=1}^\infty v_{n,n}(p_{\rm T}^{a},p_{\rm T}^{b}) \cos n\Delta\phi.
\label{eq6}
\end{eqnarray}
Here,  $v_{n,n}$ is the two-particle flow coefficient. Each one dimensional correlation function (Eq.~\ref{eq5}) is then expanded into a Fourier series according to the above equation (Eq.~\ref{eq6}), given as:
\begin{eqnarray}
C(\Delta\phi) \propto [1+ \sum_{n} 2 v_{n,n} (p_{\rm T}^{a},p_{\rm T}^{b}) cos(n\Delta\phi)].
\label{eq6.1}
\end{eqnarray}

Now, $v_{n,n}$ is calculated directly via a discrete Fourier transformation, given as
\begin{eqnarray}
v_{n,n} (p_{\rm T}^{a},p_{\rm T}^{b}) &=& \langle cos(n\Delta\phi) \rangle \nonumber\\
 &=& \frac{\sum_{m=1}^{N} cos(n\Delta \phi_m) \times C(\Delta \phi_m)}{\sum_{m=1}^{N} C(\Delta \phi_m)},
\label{eq6.2}
\end{eqnarray}
where, $N$ = 200 is the number of $\Delta\phi$ bins in the range $-\pi/2 < \Delta \phi < 3\pi/2$. Here, $v_{n,n}$ are symmetric functions with respect to $p_{\rm T}^{a}$ and $p_{\rm T}^{b}$. The definition of harmonics given in Eq.~\ref{eq2} also enters to Eq.~\ref{eq6}, which can be written as;
\begin{eqnarray}
\frac{dN_{\rm pairs}}{d\Delta\phi} \propto 1+2\sum_{n=1}^\infty v_{n}(p_{\rm T}^{a}) v_{n}(p_{\rm T}^{b}) \cos n\Delta\phi.
\label{eq7}
\end{eqnarray}
By this definition, the event plane angle drops out in convolution. So, if
the azimuthal anisotropy is driven by collective expansion then $v_{n,n}$ should factorize into the product of two single-particle harmonic coefficients~\cite{ATLAS:2012at,Chatrchyan:2013kba,Aamodt:2011by}. 
\begin{eqnarray}
v_{n,n}(p_{\rm T}^{a},p_{\rm T}^{b})= v_{n}(p_{\rm T}^{a}) v_{n}(p_{\rm T}^{b}).
\label{eq8}
\end{eqnarray}

From the above equation, $v_{n}$ is calculated as,
\begin{eqnarray}
v_{n}(p_{\rm T}^{a})= v_{n,n}(p_{\rm T}^{a},p_{\rm T}^{b})/\sqrt{v_{n,n}(p_{\rm T}^{b},p_{\rm T}^{b})}.
\label{eq9}
\end{eqnarray}

As mentioned before, for the calculation of elliptic flow one needs to remove the residual non-flow effects. This can be achieved with a proper selection of pseudorapidity gap between particle pairs. For this reason, we have compared our results with the predictions from PYTHIA8 (Angantyr) model. Figure~20 of Ref.~\cite{Bierlich:2018xfw} shows that the measurement of elliptic flow coefficient without any pseudo-rapidity gap are significantly affected by non-flow effects. With pseudorapidity gap $|\Delta\eta| > 1$, one can reduce around 20\% of non-flow contribution, but the presence of these residual non-flow effects are still significant which may affect our study based on event-shape. Thus, we use even higher pseudorapidity gap of $2<|\Delta\eta|<4.8$ as used in Ref.~\cite{Aad:2015lwa} by ATLAS collaboration. For consistency with Ref.~\cite{Aad:2015lwa}, we have also chosen similar charged-particles' transverse momentum interval of 0.5 to 5 GeV/$c$ and pseudorapidity window of $|\eta|<2.5$ to ensure a wider pseudorapidity gap of particle pairs. Figure~\ref{v22_modelcomp} shows two particle elliptic flow co-efficient ($v_{2,2}(p_{\rm T}^a,p_{\rm T}^b)$) as a function of $p_{\rm T}^a$ for low-$S_{0}$ (red triangles), high-$S_{0}$ (blue squares) and spherocity-integrated (black circles) events in minimum bias Pb-Pb collisions at $\sqrt{s_{\rm NN}} = 5.02$ TeV in PYTHIA (full markers) and AMPT (empty markers). Since Angantyr in PYTHIA8 produces a full final state, it allows for the calculation of elliptic flow coefficients, even in the absence of collective effects, giving an estimation of the presence of non-flow effects. The comparison with the results from AMPT with PYTHIA (Angantyr) will give an estimation of remaining non-flow contribution in the AMPT results and this comparison is very essential for the event shape studies as by construction the low-$S_{0}$ events may have significant non-flow effects. The comparison in Fig.~\ref{v22_modelcomp} shows that the non-flow effects with pseudorapidity gap of $2<|\Delta\eta|<4.8$ is quite negligible and it does not affect the event shape studies. Due to the absence of collective effects in PYTHIA (Angantyr), the studies based on spherocity do not show any variation in two particle elliptic flow co-efficient for low-$S_{0}$ and high-$S_{0}$ events. However, as AMPT includes collective effects, one can clearly see the variation of two particle elliptic flow co-efficient for low-$S_{0}$ and high-$S_{0}$ events. Keeping above methods in mind, we now proceed for the estimation of one dimensional azimuthal correlation, $v_{2,2}(p_{\rm T}^a,p_{\rm T}^b)$ and $v_{2}(p_{\rm T}^a)$ for different spherocity and centrality classes in Pb-Pb collisions at $\sqrt{s_{\rm NN}} = 5.02$ TeV.

\begin{figure}[ht!]
\includegraphics[scale=0.40]{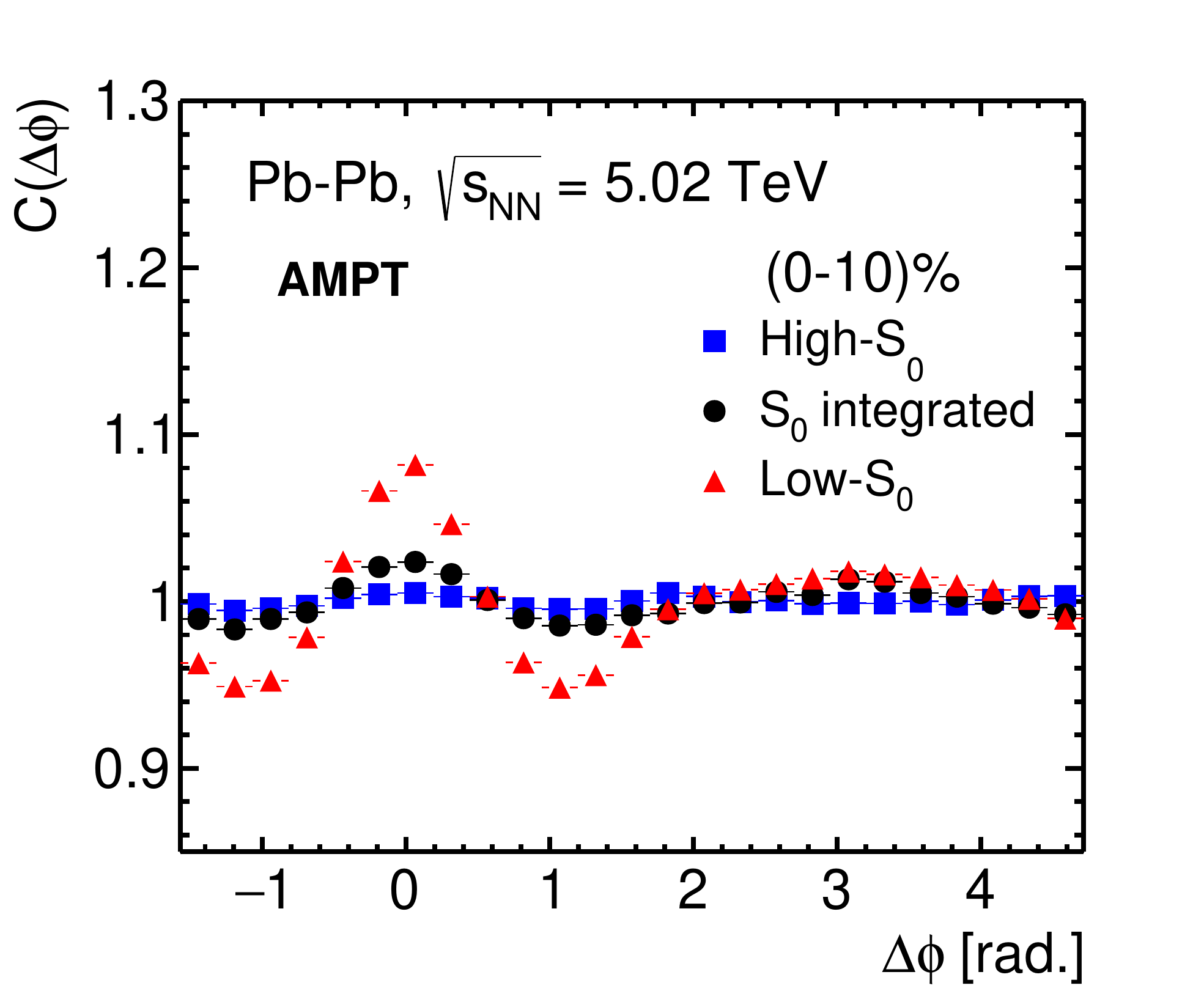}
\includegraphics[scale=0.40]{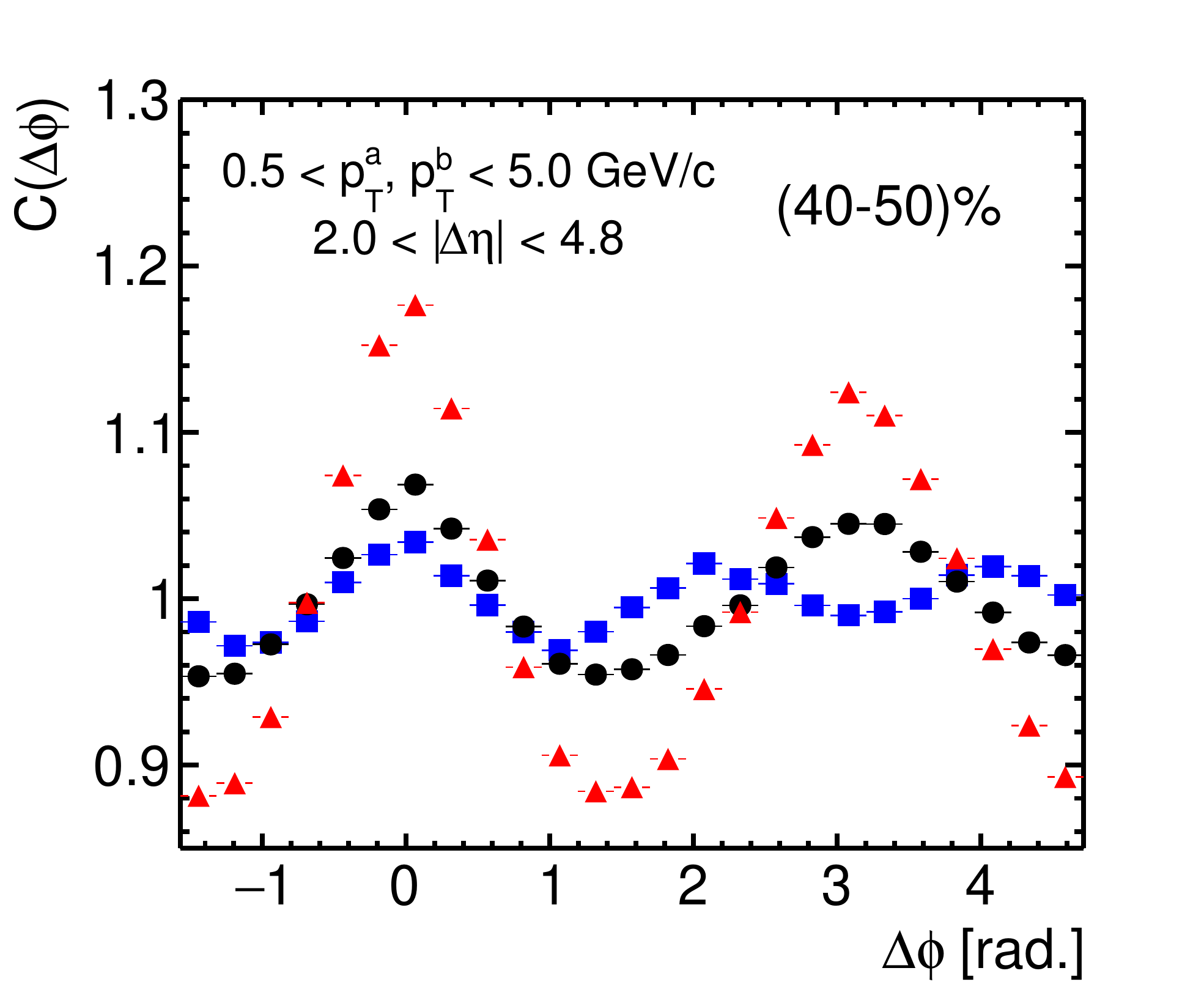}
\includegraphics[scale=0.40]{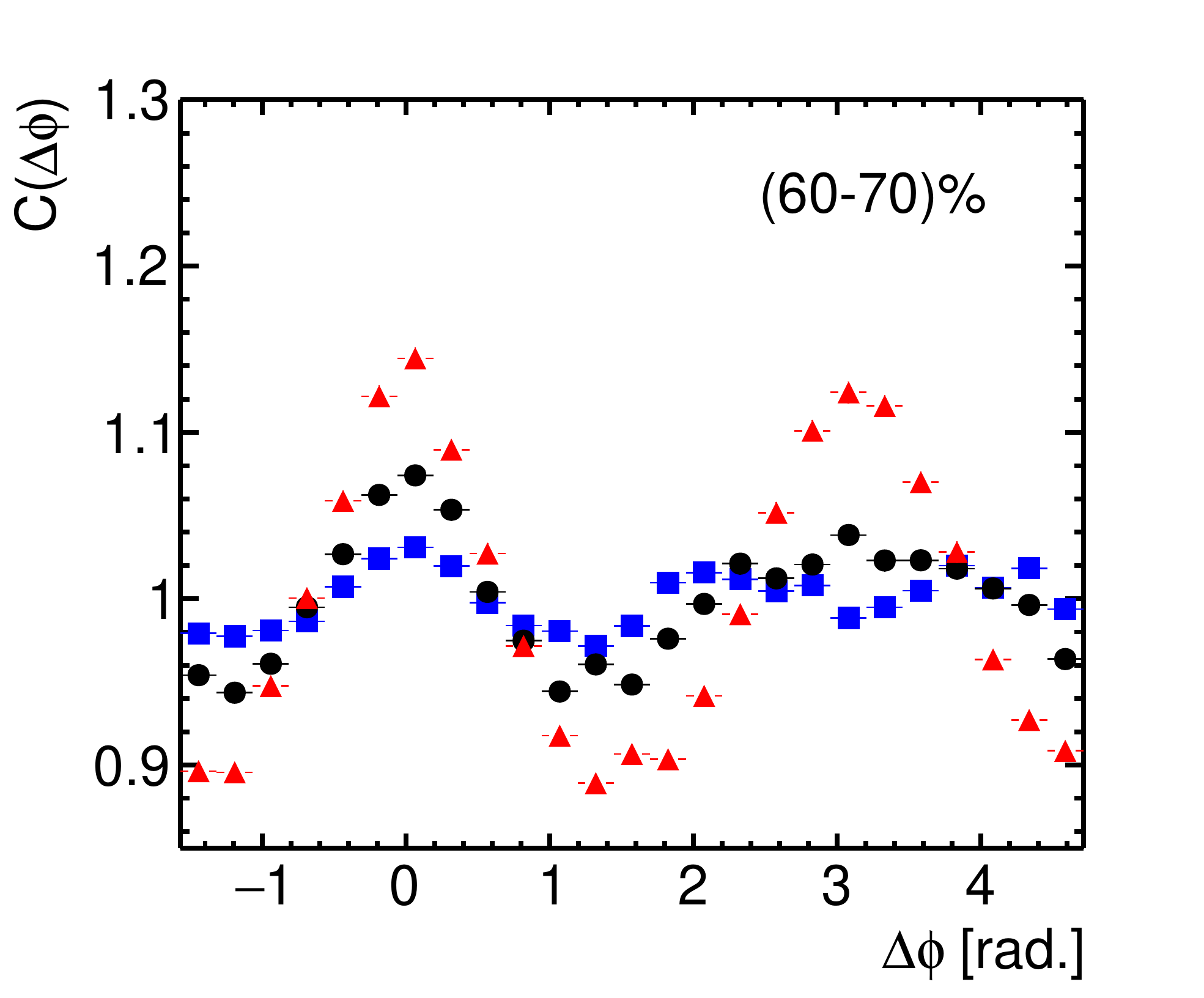}
\caption[width=18cm]{(Color Online) One dimensional azimuthal correlation of charged particles for low-$S_{0}$ (red triangles), high-$S_{0}$ (blue squares) and spherocity-integrated (black circles) events in Pb-Pb collisions at $\sqrt{s_{\rm NN}} = 5.02$ TeV for 0-10\% (top), 40-50\% (middle) and 60-70\% (bottom) centrality classes using AMPT model.}
\label{CDeltaPhi}
\end{figure}

Figure~\ref{CDeltaPhi} shows one dimensional azimuthal correlation of charged particles for low-$S_{0}$ (red triangles), high-$S_{0}$ (blue squares) and spherocity-integrated (black circles) events in Pb-Pb collisions at $\sqrt{s_{\rm NN}} = 5.02$ TeV for 0-10\% (top), 40-50\% (middle) and 60-70\% (bottom) centrality classes using AMPT model. The one dimensional azimuthal correlation functions are calculated for $2<|\Delta\eta|<4.8$ in the transverse momentum range of the particle pairs from 0.5 to 5 GeV/$c$. Figure~\ref{CDeltaPhi} suggests that the magnitude of the modulation correlates strongly with spherocity, which reflects the fact that using spherocity one can differentiate events based on geometrical shapes. From the comparison of 1D correlation plots across centrality classes, one can observe that the correlation is stronger in semi-central collisions compared to most peripheral and most central collisions. While going from central to peripheral collisions, double peaks for away side ($\Delta\phi \sim \pi$) appears for high-$S_{0}$ events. This structure may reflect the residual contribution of the triangular flow~\cite{Aad:2015lwa}. The behavior seen in Figure~\ref{CDeltaPhi} is analogous to the behavior observed in Ref.~\cite{Aad:2015lwa}, where flow vector was used to study the event shape dependence.

\begin{figure}[ht!]
\includegraphics[scale=0.4]{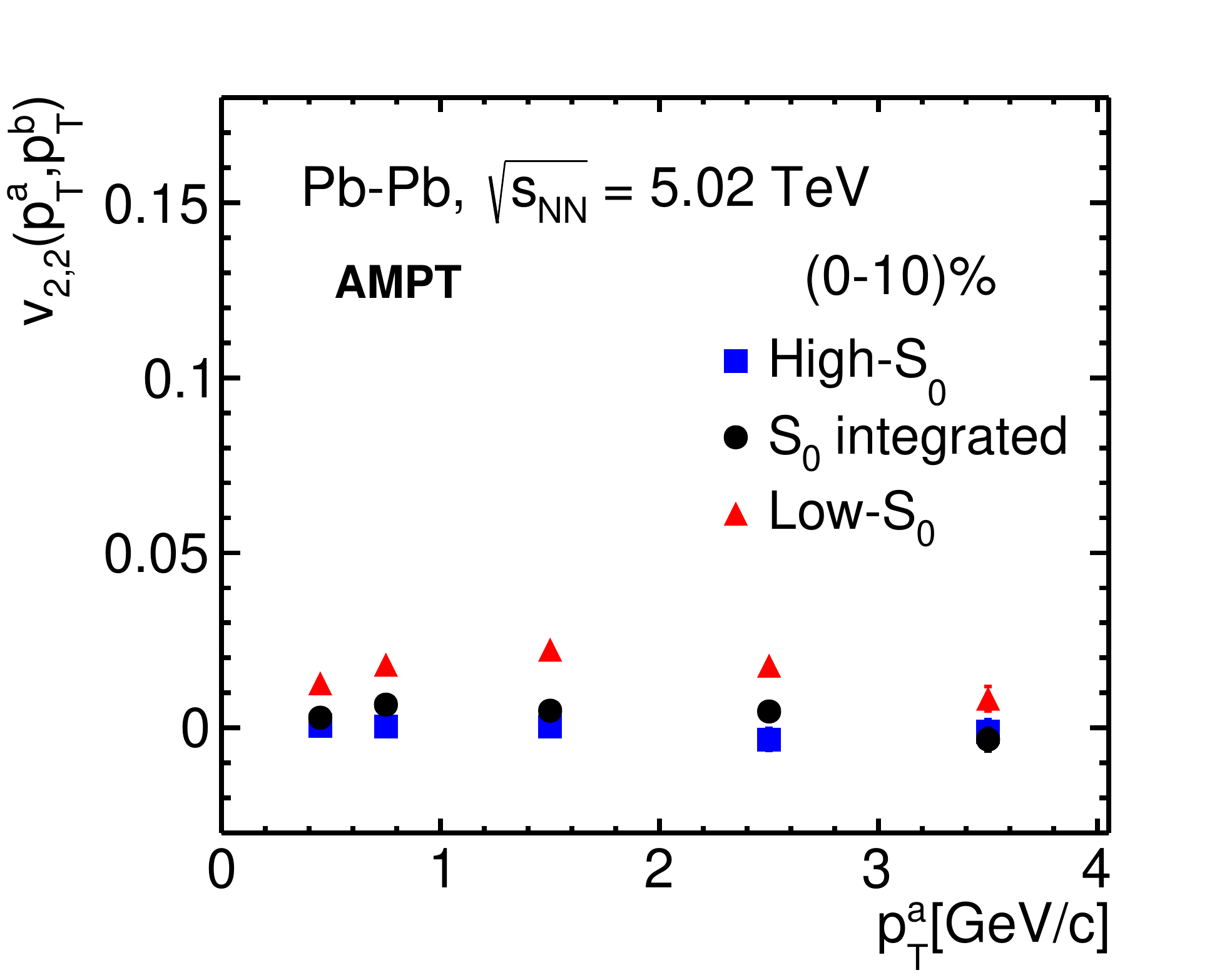}
\includegraphics[scale=0.4]{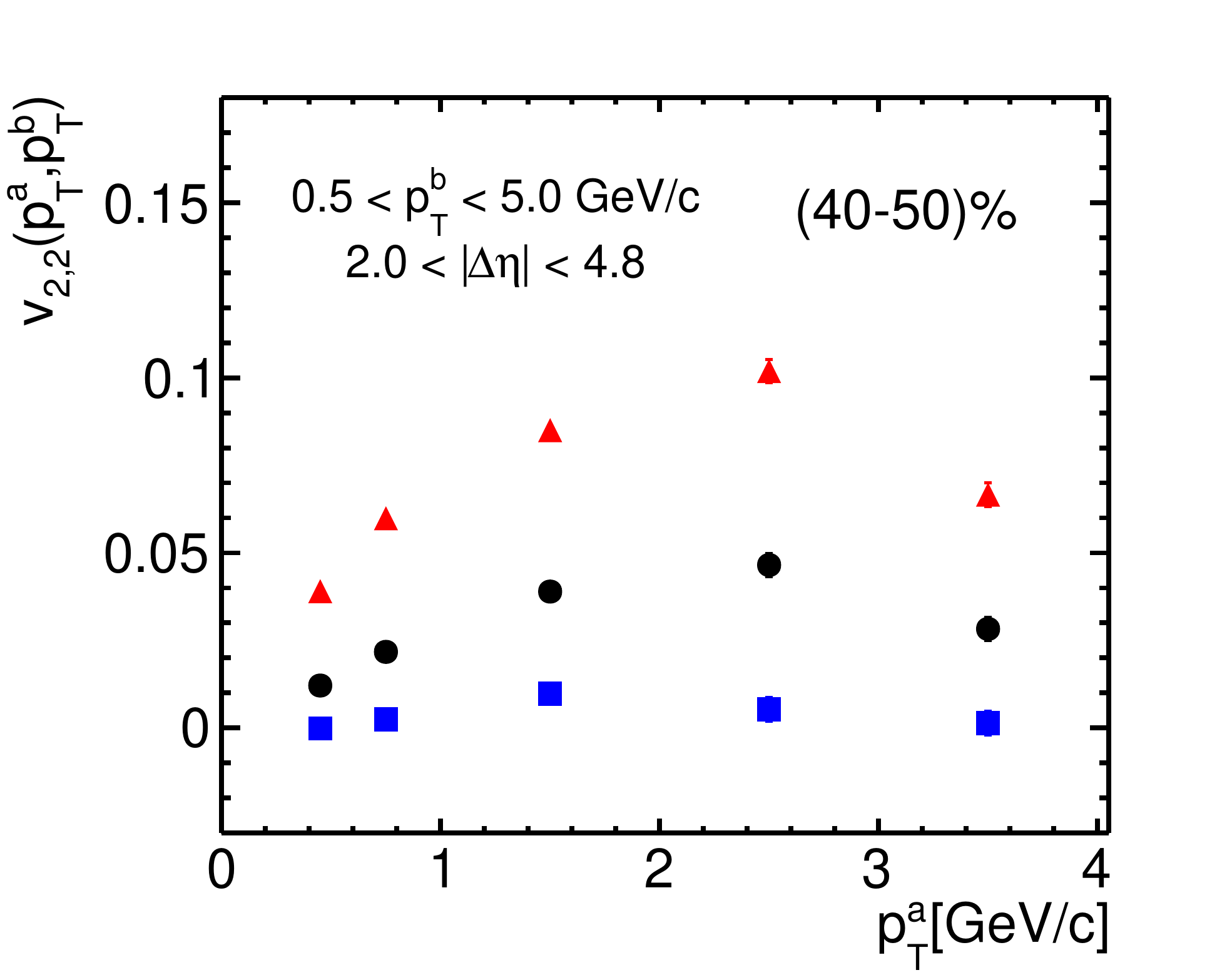}
\includegraphics[scale=0.4]{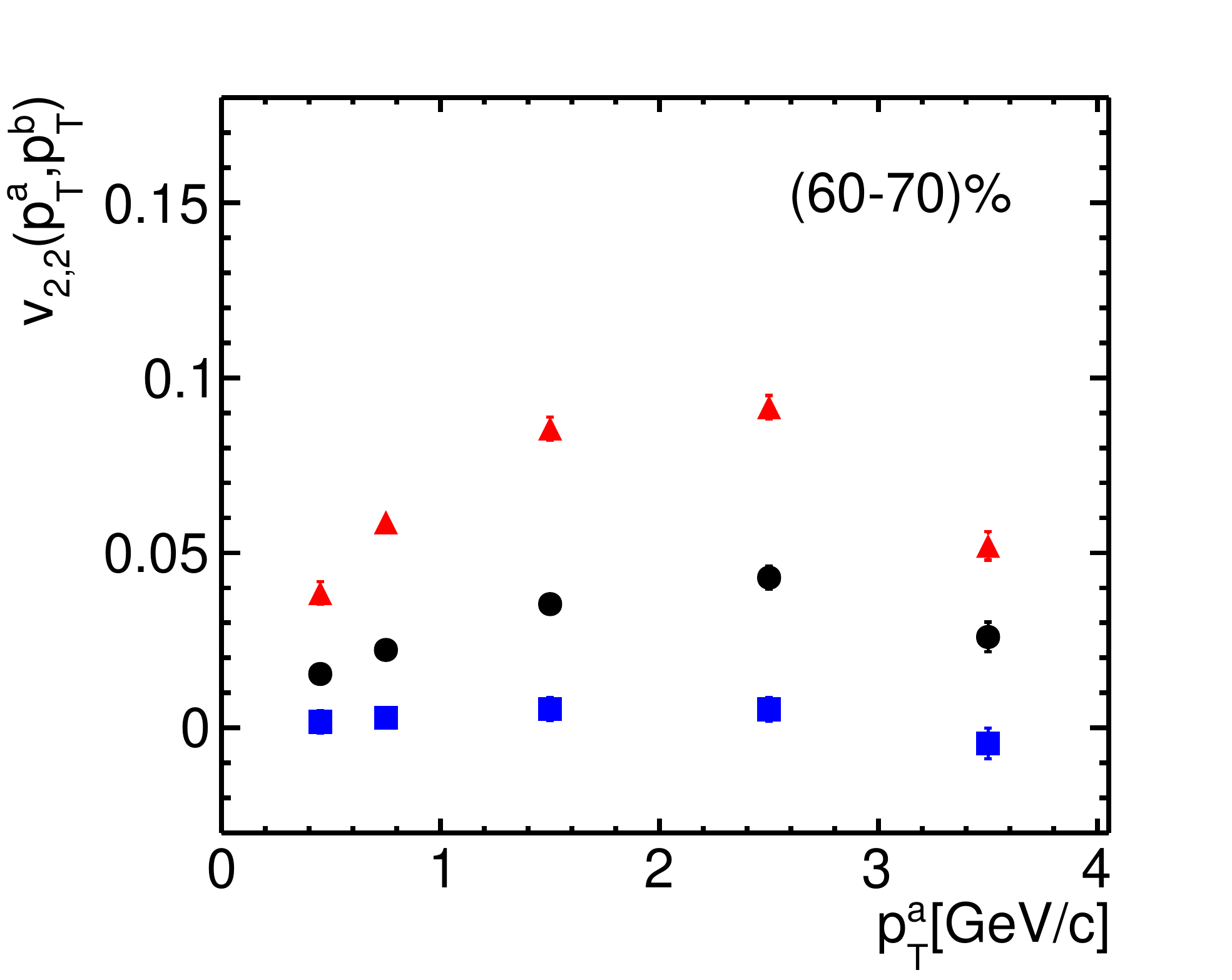}
\caption[width=18cm]{(Color Online) Two particle elliptic flow co-efficient ($v_{2,2}(p_{\rm T}^a,p_{\rm T}^b)$) of charged particles as a function of $p_{\rm T}^a$ of charged particles for low-$S_{0}$ (red triangles), high-$S_{0}$ (blue squares) and spherocity-integrated (black circles) events in Pb-Pb collisions at $\sqrt{s_{\rm NN}} = 5.02$ TeV for 0-10\% (top), 40-50\% (middle) and 60-70\% (bottom) centrality classes using AMPT model.}
\label{v22_pTa}
\end{figure}

\begin{figure}[ht!]
\includegraphics[scale=0.4]{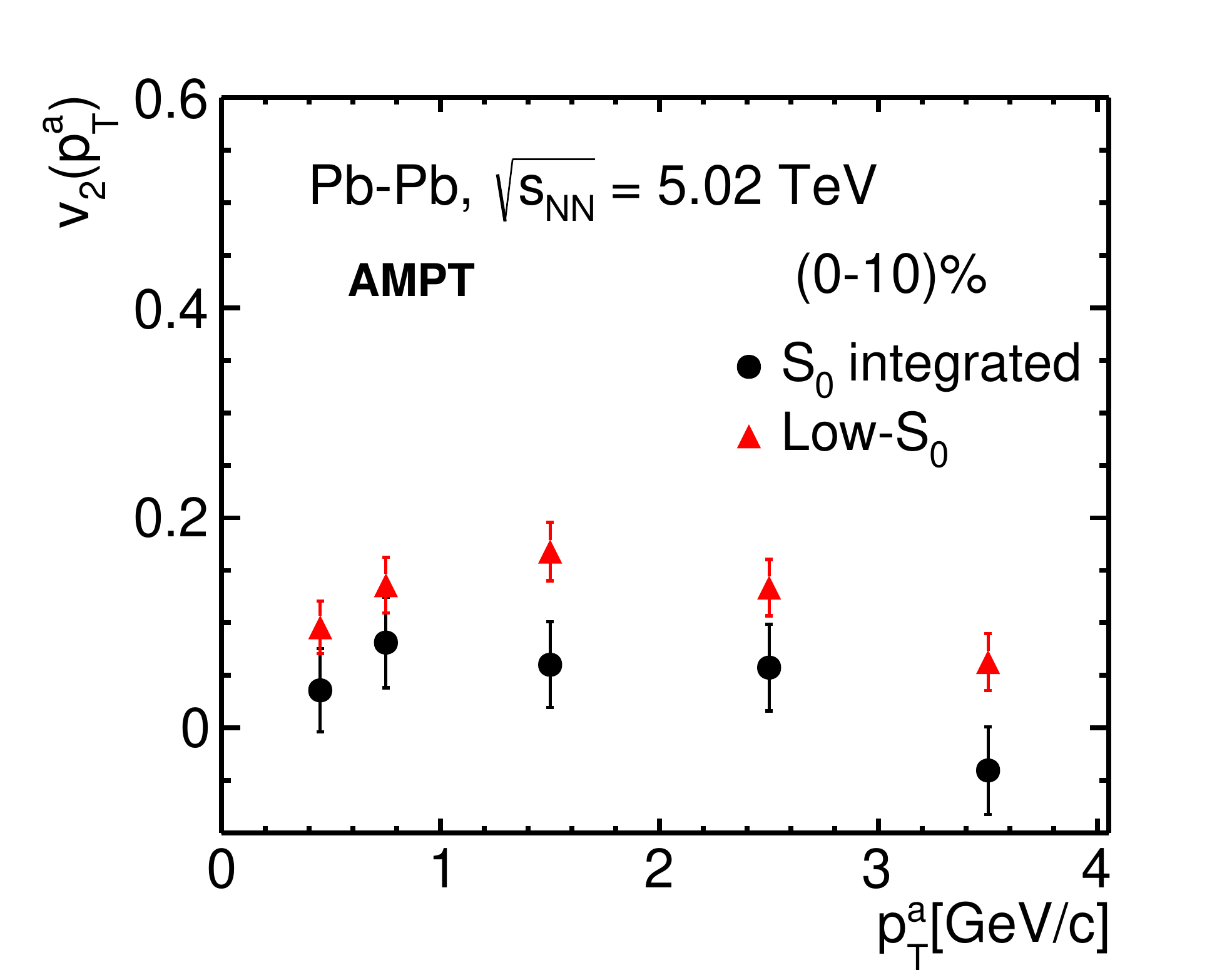}
\includegraphics[scale=0.4]{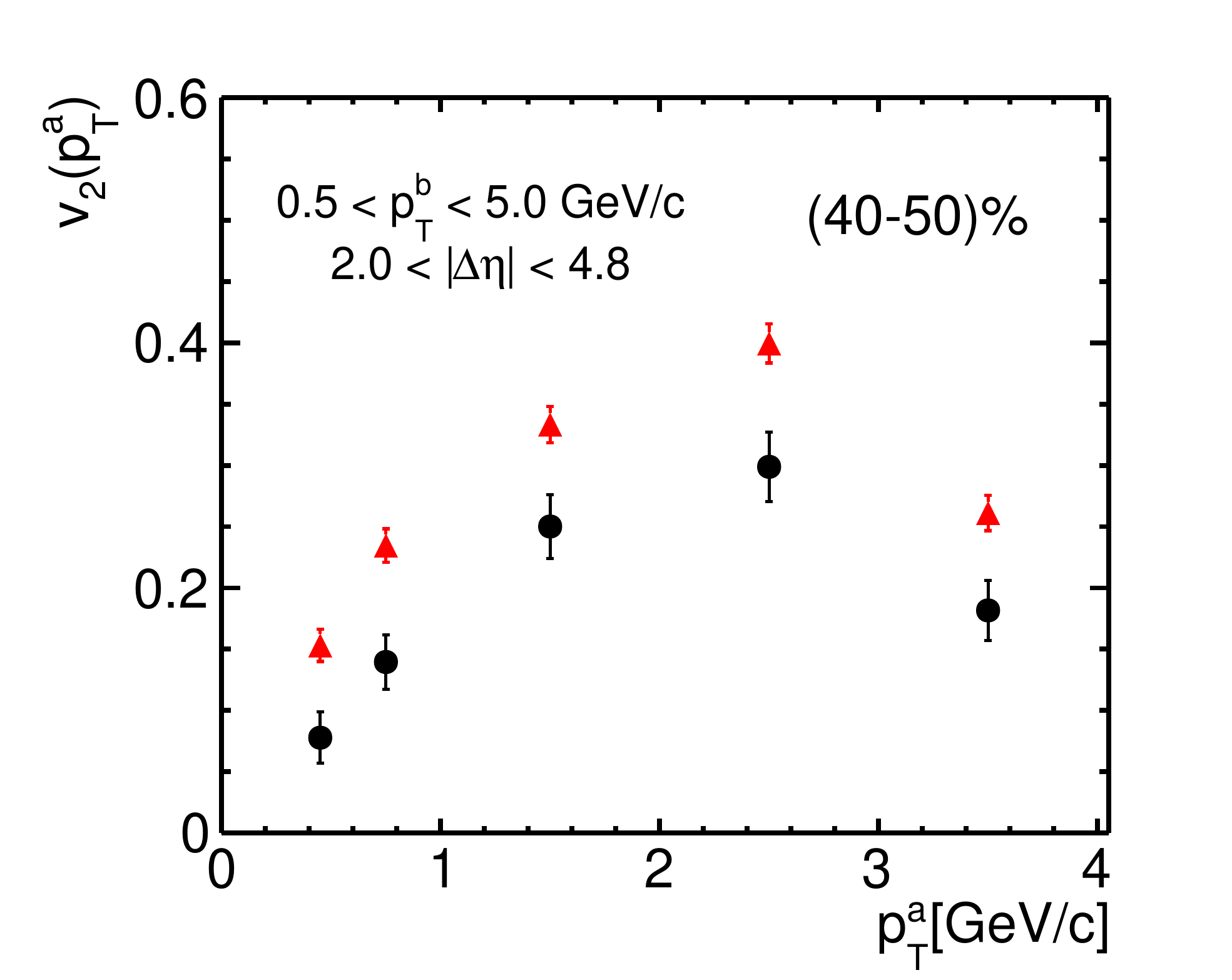}
\includegraphics[scale=0.4]{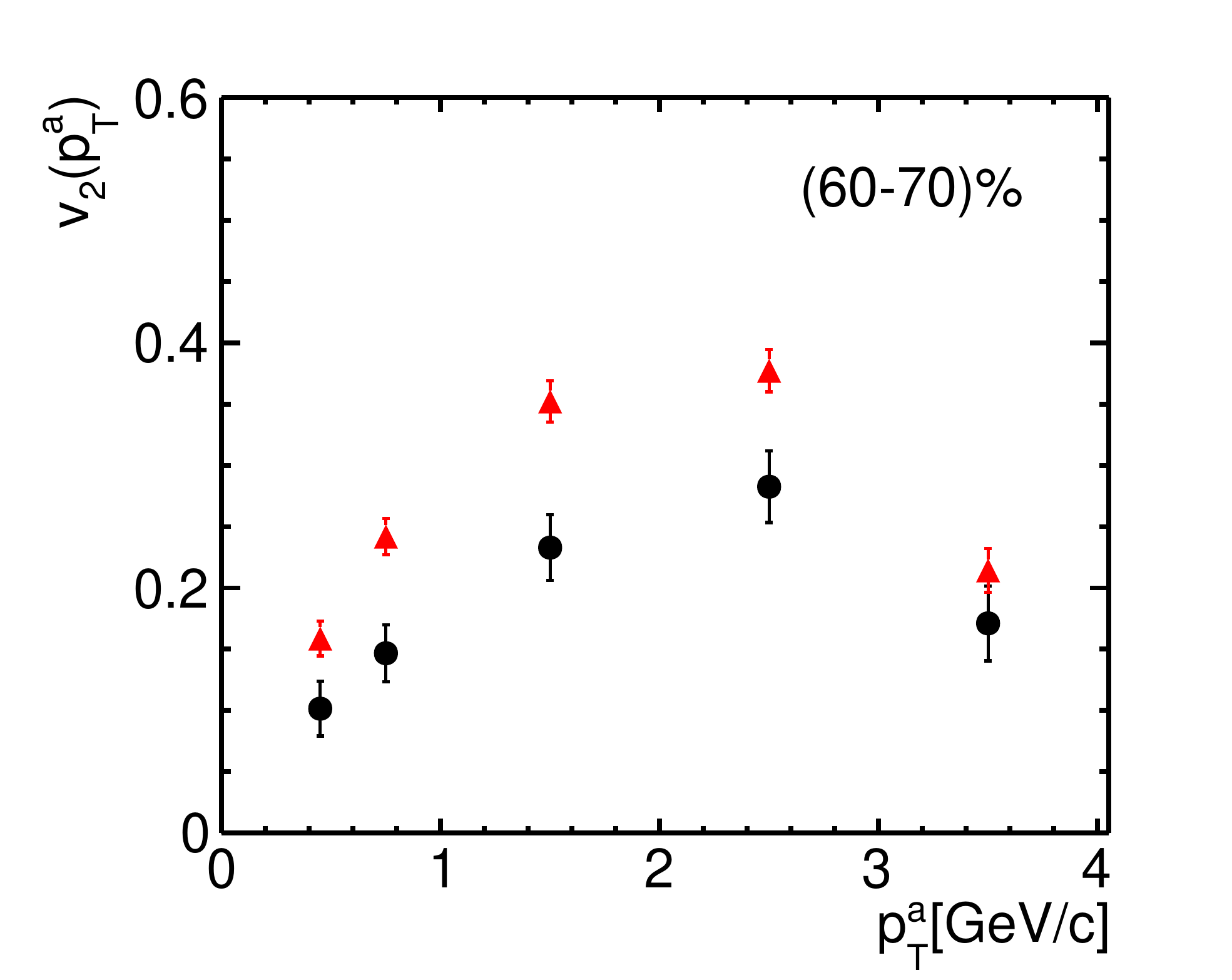}
\caption[width=18cm]{(Color Online) Single particle elliptic flow co-efficient ($v_{2}(p_{\rm T}^a)$) of charged particles as a function of $p_{\rm T}^a$ for low-$S_{0}$ (red triangles) and spherocity-integrated (black circles) events in Pb-Pb collisions at $\sqrt{s_{\rm NN}} = 5.02$ TeV for 0-10\% (top), 40-50\% (middle) and 60-70\% (bottom) centrality classes using AMPT model.}
\label{v2_pTa}
\end{figure}

Figures~\ref{v22_pTa} and \ref{v2_pTa} show two particle ($v_{2,2}(p_{\rm T}^a,p_{\rm T}^b)$) and single particle ($v_{2}(p_{\rm T}^a)$) elliptic flow co-efficients of charged particles as a function of $p_{\rm T}^a$, respectively for different spherocity classes in Pb-Pb collisions at $\sqrt{s_{\rm NN}} = 5.02$ TeV for 0-10\% (top), 40-50\% (middle) and 60-70\% (bottom) centrality classes using AMPT model. The elliptic flow coefficients are calculated for pseudo-rapidity gap of $2<|\Delta\eta|<4.8$ in the transverse momentum range of the particle pairs from 0.5 to 5 GeV/$c$. As we move from 0-10\% to 40-50\% central collisions, two particle and single particle elliptic flow co-efficient for spherocity-integrated events increase and become maximum for mid-central collisions. However, the values slightly decrease if we move further towards peripheral collisions. For most central collisions, the system has less initial spatial anisotropy, thus it has less value of $v_2$. In most peripheral collisions, the nuclear overlap region at the collision point decreases and the size as well as the density of participating partons also decrease. So, less number of particles emerge from this type of collisions can not carry the effect of $v_2$ till the final state. But, in case of mid-central collisions, the system has finite spatial anisotropy and the nuclear overlap region has enough participants, hence the produced system has the maximum $v_2$. The qualitative centrality dependent trend of $v_2$ is similar with the experimental results discussed in Refs.~\cite{Adam:2016izf,Aad:2015lwa,Acharya:2018ihu}.

Now, coming to the different event types based on transverse spherocity, we observe that the contribution towards $v_{2,2}$ is mostly dominated by the low-$S_{0}$ events. It can be seen in Fig.~\ref{v22_pTa}. An interesting point to notice is that, high-$S_{0}$ events have almost zero $v_{2,2}$. This clearly shows that the types of events have important role towards the initial state anisotropy in the system. high-$S_{0}$ events showing less $v_{2,2}$ is a testimony of transverse spherocity successfully separating low-$S_{0}$ and high-$S_{0}$ events through proper event topological selections. The process of isotropization, resulting in events with higher probability of high-$S_{0}$ events diminishes the azimuthal anisotropy in the final state. Figure~\ref{v2_pTa} shows the $v_2$ estimated from $v_{2,2}$ for low-$S_{0}$ and spherocity-integrated events. As the high-$S_{0}$ events have nearly zero $v_{2,2}$, it becomes nearly impossible to calculate the $v_2$ using the Eq.~\ref{eq9} as the denominator becomes nearly zero. Figure~\ref{v2_pTa} clearly indicates that the low-$S_{0}$ events dominate the elliptic flow contribution even after the removal of residual non-flow effects. These results along with the preliminary measurements by ALICE collaboration at the LHC~\cite{adrianQM} suggests that using spherocity one can study the particle production in both heavy-ion and pp collisions with varying degree of collective effects. The results based on spherocity also complements the analysis based on flow vector~\cite{Aad:2015lwa}.

We have also compared the AMPT results for spherocity-integrated events with experimental data, which can be found in the Appendix section. It would be very interesting to see how our results as a function of spherocity compare with future experimental data. We believe that the current study will act as a very nice baseline for future event shape dependence experimental work in heavy-ion collisions.

\section{Summary}
\label{section4}
In summary, we report the first implementation of transverse spherocity analysis for Pb-Pb collisions at $\sqrt{s_{\rm NN}} = 5.02$ TeV at the Large Hadron Collider using A Multi-Phase Transport Model (AMPT). The results show that transverse spherocity successfully differentiates the heavy-ion collisions' event topology based on their geometrical shapes i.e. high-$S_{0}$ and low-$S_{0}$. The comparison of the predictions from AMPT and PYTHIA8 (Angantyr) suggests that our results of elliptic flow are almost free from the residual non-flow effects. The elliptic flow as a function of transverse spherocity shows that the high-$S_{0}$ events have nearly zero elliptic flow while low-$S_{0}$ events contribute significantly to elliptic flow of spherocity-integrated events. Thus, using transverse spherocity one can classify the events based on different degree of the collective effects in heavy-ion collisions at the LHC. We believe that the results are very encouraging and an experimental exploration in this direction would be highly helpful to understand event topology dependence of system dynamics in heavy-ion collisions. 

\section*{Acknowledgements}
 One of the authors, R.S. acknowledges the financial supports from DAE-BRNS Project No. 58/14/29/2019-BRNS. The authors would like to acknowledge the usage of resources  of the LHC grid computing facility at VECC, Kolkata and computing farm at ICN-UNAM. A.O. and S.T. acknowledge the financial support from CONACyT under the Grant No. A1-S-22917. S.T. acknowledges the support from the postdoctoral fellowship of DGAPA UNAM.
 
 \section*{Appendix}
 
 \begin{figure}[ht!]
\centering
\includegraphics[scale=0.40]{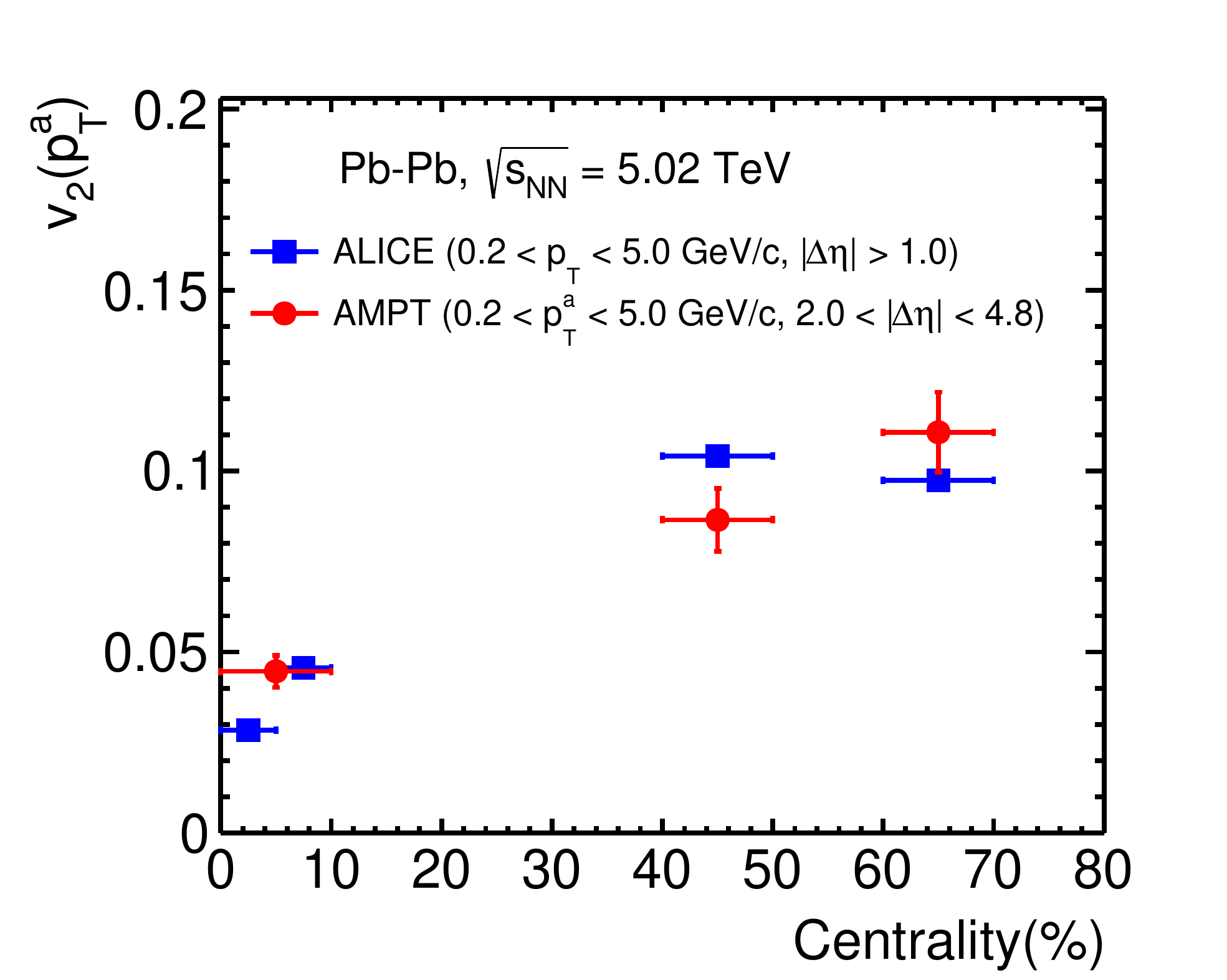}
\caption[width=18cm]{(Color Online) Comparison of $v_{2}(p_{\rm T}^a)$ of charged particles as a function of centrality in Pb-Pb collisions at $\sqrt{s_{\rm NN}} = 5.02$ TeV from ALICE~\cite{Adam:2016izf} (blue squares) and AMPT model (red circles).}
\label{v2_datacomp}
\end{figure}
 
As, the elliptic flow results from experiments at similar pseudorapidity gap for Pb-Pb collisions at $\sqrt{s_{\rm NN}} = 5.02$ TeV are currently not available, we have compared the results of spherocity-integrated events from AMPT with the published results with ALICE at the LHC~\cite{Adam:2016izf} measured in a pseudorapidity gap of $|\Delta\eta| > 1$. So, ALICE results may have slightly higher non-flow effects than the AMPT results~\cite{Bierlich:2018xfw}. Also, the event selection is a bit different in the current study due to the fact that for the calculation of spherocity, the events are chosen with at least five charged particles while no such cuts were used for the event selection in ALICE. Due to the above-mentioned reasons, one may expect slightly different values of elliptic flow in our case compared to the ALICE results. Figure~\ref{v2_datacomp} shows the comparison of AMPT predictions and ALICE results, which indicates that default settings of SM mode of AMPT gives a good prediction for the experimental data within uncertainties. To accurately match with the experimental data, one can vary the tunes of the AMPT model, which is out of the scope of this manuscript.


\end{document}